\documentclass[
reprint,
preprintnumbers,
superscriptaddress,
nofootinbib,
amsmath,amssymb,
aps,
prd,
]{revtex4-2}

\usepackage[caption=false]{subfig}
\usepackage{graphicx}
\usepackage[usenames,dvipsnames]{xcolor} 
\usepackage[utf8]{inputenc}
\usepackage{color}
\usepackage{hyperref}
\usepackage[normalem]{ulem} 
\usepackage{physics}

\usepackage{enumitem}
\usepackage{comment}
\usepackage{bm}

\graphicspath{{./}}

\hypersetup{
  colorlinks   = true, 
  urlcolor     = blue, 
  linkcolor    = blue, 
  citecolor   = blue 
}

\begin{document}
\title{Constraining the neutron star-black hole
merger rate}
\author{Ian Harry}
\affiliation{Institute of Cosmology and Gravitation, University of Portsmouth, Portsmouth, PO1 3FX, UK}
\author{Charlie Hoy}
\email{charlie.hoy@port.ac.uk}
\affiliation{Institute of Cosmology and Gravitation, University of Portsmouth, Portsmouth, PO1 3FX, UK}

\begin{abstract}

Current template-based gravitational-wave searches for compact binary mergers neglect the general relativistic phenomenon of spin-induced orbital precession. Owing to their asymmetric masses, gravitational-waves from neutron star-black hole (NSBH) binaries are prime candidates for displaying strong imprints of spin-precession. Current searches may therefore miss a significant fraction of the astrophysical population, and the detected NSBH population may be significantly suppressed or biased. Here we report the most sensitive search for NSBH binaries to date by including spin-precession for the first time. We analyze data from the entirety of the third LIGO--Virgo--KAGRA gravitational-wave observing run and show that when accounting for spin-precession, our search is up to 100\% more sensitive than the search techniques currently adopted by the LIGO--Virgo--KAGRA collaboration (for systems with strong precessional effects). This allows us to more
tightly constrain the rate of NSBH mergers in the local Universe. When focusing on a potentially precessing subpopulation of NSBH mergers, the lack of observed candidates allows us to place an upper limit on the merger rate of $R_{90} = 79\, \mathrm{Gpc}^{-3}\mathrm{yr}^{-1}$ with 90\% confidence. We then show that if there is no preferred direction of component spin, the overall rate of NSBH mergers is on average 16\% smaller than previously believed. Finally, we report four new subthreshold NSBH candidates, all with strong imprints of spin precession, but note that these are most likely to be of terrestrial origin.
\end{abstract}

\maketitle

\textbf{\textit{Introduction.}}— The standard approach to observe gravitational-wave (GW) signals from neutron star-black hole (NSBH) binaries is through template-based matched filter searches~\cite{Usman:2015kfa,Sachdev:2019vvd,Chu:2020pjv,Aubin:2020goo}, a well established signal processing technique. Here, a GW signal masked by noise is identified by matching the data against a pregenerated bank of templates. A limitation of template-based matched filter searches is that signals may remain undetected if they are not sufficiently similar to at least one template in the bank. For GW astronomy, templates are constructed from theoretical models, and to date LIGO--Virgo--KAGRA (LVK) collaboration analyses are restricted to templates that neglect the general-relativistic phenomenon of spin-induced orbital precession~\cite{Apostolatos:1994mx}. Spin-precession arises when there is a non-zero effective perpendicular spin~\cite{Schmidt:2014iyl}, and hence a spin angular momentum that is mis-aligned with the orbital angular momentum of the binary. A binary with significant spin-precession has effective perpendicular spin close to unity, while a binary with spins aligned with the orbital angular momentum has effective perpendicular spin close to zero.

Restricting templates to ignore spin-precession implies that a significant fraction of the NSBH population may remain undetected~\cite{DalCanton:2014qjd,Harry:2016ijz}. If we assume that current template banks can only detect NSBH binaries with minimal spin-precession (effective perpendicular spin $< 0.1$), we estimate that we could be missing as many as $\sim 85\%$ of GWs from NSBH binaries formed in the local Universe (assuming an agnostic population of NSBHs, with the neutron star's spin $< 0.05$ inline with observations from Galactic binaries~\cite{Burgay:2003jj}). This may impact our ability to accurately infer the NSBH merger rate from GW observations; a more sensitive search may not only uncover more GW signals, but may also explore a greater spacetime volume $\langle VT\rangle$.

Current search pipelines are tailored to detect GWs from binaries formed through the isolated channel. This is because minimal spin-precession is expected: binaries formed through the isolated channel evolve from a pair of massive binary stars, which shed angular momentum through stable and unstable mass transfer. Although natal kicks generated through core collapse can cause small mis-alignments, we expect the spins of isolated binaries to be approximately aligned with the orbital angular momentum~\cite{Kalogera:1999tq, Chattopadhyay:2020lff, Broekgaarden:2021iew}. Current search pipelines are likely to miss GWs from binaries formed through the dynamic channel -- binaries formed in a dense stellar environment such as globular clusters. The reason is because these binaries are believed to have an isotropic distribution of spins, meaning that spin-precession is more likely~\cite{Rodriguez:2016vmx}. The recent observation that the NSBH GW200105\_162426~\cite{LIGOScientific:2021qlt}
may have non-negligible eccentricity~\cite{Morras:2025xfu}---which would indicate that it most
likely formed through the dynamical channel---adds further weight to
the possibility of a population of dynamically formed NSBH binaries.
It is of great astrophysical interest to be able to identify, or equally rule out, NSBHs with spin-precession as it will allow the formation mechanism of NSBHs in the Universe to be determined.

In this \emph{Letter}, we present for the first time a dedicated template-based matched filter search for NSBHs which is sensitive to the whole NSBH population. By restricting attention to data from the third GW observing run~\cite{KAGRA:2023pio}, O3, we find that our search is on average $44\%$ more sensitive than the search techniques currently adopted by the LVK,
with sensitivity most improved (up to 100\%) for systems demonstrating strong imprints of spin-precession. This enables us to place 
tighter constraints on the rate of NSBH mergers in the local Universe.
We report an upper limit on the merger rate of a precessing subpopulation
of NSBH mergers to be $R_{90} = 79\, \mathrm{Gpc}^{-3}\mathrm{yr}^{-1}$ with 90\% confidence. We then further constrain the overall rate
of NSBH mergers to be on average 16\% smaller than the rate we compute assuming a search that requires no misalignment between component spin and orbital angular momentum (similar to those currently adopted by the LVK). Being able to observe and correctly quantify the NSBH merger rate is of significant astrophysical interest as it will allow their underlying source parameters to be measured more accurately. This enables (for example) a better understanding of the impact of supernova kicks~\cite{Vitale:2014mka}. This letter is accompanied by~\cite{Harry:2025aaa} (hereafter Supplemental Material) which includes supplementary information on our matched filter search, and how we inferred the merger rate of NSBH binaries in the local Universe. Supplemental Material~\cite{Harry:2025aaa} includes Refs.~\cite{Allen:2005fk,Babak:2012zx, Usman:2015kfa, Sachdev:2019vvd, Venumadhav:2019tad, Chu:2020pjv, Aubin:2020goo,Harry:2016ijz,McIsaac:2023ijd,Harry:2016ijz,Schmidt:2024jbp,Fairhurst:2019vut,Pratten:2020ceb,Pratten:2020fqn,Schmidt:2012rh,Hamilton:2021pkf,Dhurkunde:2022aek,LIGOScientific:2016ebi,metropolis1949monte,KAGRA:2023pio,Veitch:2014wba,Skilling2004,Skilling:2006gxv,Speagle:2019ivv,Ashton:2018jfp,Romero-Shaw:2020owr,KAGRA:2021vkt,Morisaki:2021ngj,Thompson:2020nei,Matas:2020wab,Gonzalez:2022prs,LIGOScientific:2024elc,Kumar:2016zlj,Cornish:2014kda,Farr:calnote,ligo_scientific_collaboration_and_virgo_2021_5546663,Mandel:2018mve,Farr:2019rap,Tiwari:2017ndi,2019PhRvD.100d3030T,Biscoveanu:2022iue,LIGOScientific:2021qlt}.

\textbf{\textit{Matched filter searches.}}— Template-based matched filter searches identify GW signals by matching the data against ``template'' waveforms representing predicted GW signals. A significant issue for matched-filter searches is the dimensionality of the problem. When neglecting additional terms that describe differences between neutron stars and black holes, for example the tidal deformability of neutron stars~\cite[see e.g.][]{Chatziioannou:2020pqz} which is identically zero for black holes~\cite{Damour:2009vw,Binnington:2009bb}, a NSBH merger on a circular orbit is described by fifteen parameters, and one must be able to observe the signal anywhere in this fifteen dimensional parameter space.

In current LVK searches, the dimensionality of the problem is decreased by assuming that the angular momentum of the two component bodies is aligned, or anti-aligned, with the orbit. One is then able to analytically maximise the signal-to-noise ratio (SNR) over the binary's sky location, orientation, luminosity distance, and time of coalescence, reducing the problem to 4 dimensions; the two component masses, and two component spin magnitudes. An interested reader can find detailed descriptions of current methods in~\cite{Allen:2005fk, Babak:2012zx, Usman:2015kfa, Sachdev:2019vvd, Venumadhav:2019tad, Chu:2020pjv, Aubin:2020goo}.

While undeniably successful, this technique results in a significant loss in sensitivity for signals where spins are misaligned with the orbit, in particular systems where this misalignment causes a precession of the orbital plane~\cite{Harry:2016ijz}. In~\cite{McIsaac:2023ijd} we recently proposed a novel method for incorporating misaligned spins in searches. We apply that method here to carry out a search for NSBH mergers with misaligned spins in data from the third LVK observing run. We summarize the method in Supplementary Material~\citep{Harry:2025aaa}, including details on our template bank and the generation process.

\begin{figure*}[t!]
\begin{minipage}{0.5\textwidth}
    \begin{center}
        \includegraphics[width=\textwidth]{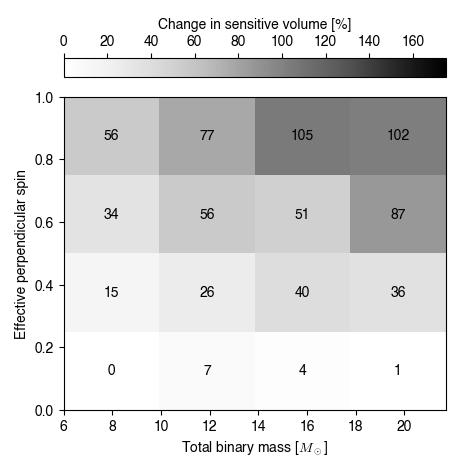}
    \end{center}
\end{minipage}
\hfill
\begin{minipage}{0.46\textwidth}
    \vspace{2em}
    \begin{center}
        \includegraphics[width=\textwidth]{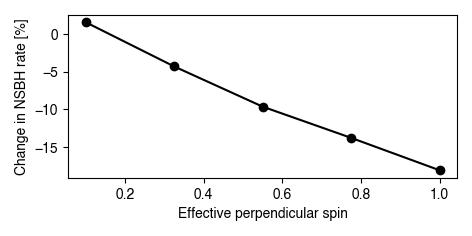}
        \includegraphics[width=\textwidth]{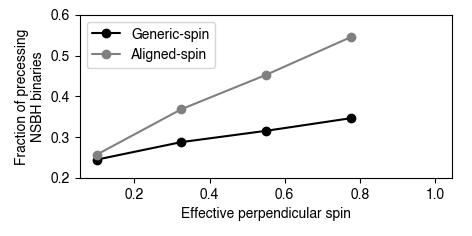}
    \end{center}
\end{minipage}
\caption{
  Percentage change in the search sensitivity to NSBH mergers (\emph{Left}) and percentage change in the merger rate of NSBHs in the local Universe (\emph{Upper Right}) between a template bank which includes spin precession (generic-spin) and a template bank which restricts spins to be aligned with the orbital angular momentum (aligned-spin). The change in sensitive volume is shown as a function of the binaries total mass and the degree of spin-precession (encoded by the effective perpendicular spin $\chi_{\mathrm{p}}$). The change in NSBH merger rate is shown just as a function of the effective perpendicular spin. Assuming the full parameter space, $\chi_{\mathrm{p}} \leq 1$, we infer the NSBH merger rate to be $16\%$ smaller than previously published results. We also show the fraction of precessing NSBH binaries in the Universe as a function of $\chi_{\mathrm{p}}$ (\emph{Lower Right}). According to our generic spin search, we estimate that no more than $30\%$ of NSBHs in the Universe undergo spin-precession for an independent subpopulation of NSBH binaries with $\chi_{\mathrm{p}} > 0.4$.
  }
\label{fig:rate}
\end{figure*}

\textbf{\textit{Estimating the neutron star-black hole merger rate.}}— In this work, we focus on constraining the NSBH merger rate. The NSBH merger rate is estimated through GW observations by combining the search sensitivity with the number of confident GW signals detected. To access a search's sensitivity, a search is performed on simulated signals drawn from a reference population model that has been added to real GW detector noise. Based on the number of signals injected and recovered, a sensitivity estimate can be calculated~\cite{Tiwari:2017ndi}, see Supplementary Material~\citep{Harry:2025aaa} for details.

The novel inclusion of spin precession in our search means that we achieve on average $44\%$ greater sensitivity to NSBH mergers compared to search techniques currently adopted by the LVK~\cite{Usman:2015kfa}. This increases to an average of $85\%$ for highly precessing NSBH systems. This can be seen in the left-hand panel of Figure~\ref{fig:rate}, which compares the sensitivity of our generic-spin search with a comparable search using only aligned-spin waveforms. For highly precessing (effective perpendicular spins $> 0.75$), and high total mass NSBH systems (total mass $> 14\, M_{\odot}$), we observe $> 100\%$ increase in the sensitive volume. Although it is possible to marginalize over multiple discrete population models when estimating the sensitive volume, we assume a single, agnostic population owing to a large uncertainty in the underlying NSBH population, see Supplementary Material~\citep{Harry:2025aaa} for details. This analysis ignores the possibility of the neutron star being tidally disrupted during merger and deformed during the inspiral. Tidal disruption is unlikely to play a pivotal role, since most NSBH binaries are likely to merge before the neutron star is disrupted for any expected neutron star equation of state~\cite{Foucart:2013psa}. Likewise, searches that ignore tidal deformability are likely to capture more than 99\% of GW signals from NSBH binaries~\cite{Pannarale:2011pk}.

Figure~\ref{fig:rate} shows the largest difference in sensitive volume for NSBH systems with large effective perpendicular spin. Although the maximum neutron star spin observed from Galactic binary neutron star mergers is $< 0.05$~\cite{Burgay:2003jj} we argue that highly precessing NSBH binaries are astrophysically possible. Given that the effective perpendicular spin is highly dependent on the dynamics of the primary object for asymmetric component mass binaries~\cite{Schmidt:2014iyl}, highly precessing NSBH systems require the black hole to have large spin magnitudes that are misaligned with the orbital angular momentum. Firstly, \emph{highly spinning black holes are possible}: although the current black hole population estimates from GW observations point towards low spin magnitudes, with half of spin magnitudes less than 0.26 (based on binary black hole mergers)~\cite{KAGRA:2021duu}, rapidly spinning black holes have been observed through GW~\cite{Hannam:2021pit,KAGRA:2021vkt} and X-ray observations~\cite{Valsecchi:2010cw,Wong:2011eg,Zhao:2021bhj}. In fact, Cygnus X-1 is expected to form an NSBH binary with a near-maximal spin black hole~\cite{Belczynski:2011bv}. Secondly, \emph{spin magnitudes misaligned with the orbital angular momentum are possible}: NSBH binaries formed through dynamic capture are expected to produce precessing binaries~\cite{Rodriguez:2016vmx}. Similarly, although binaries formed in isolation are expected to have spins primarily aligned with the orbital angular momentum~\cite{Kalogera:1999tq}, precessing binaries are possible in certain regions of the parameter space~\cite{Steinle:2020xej}, and as a consequence of supernova kicks~\cite{Rodriguez:2016vmx,Gerosa:2018wbw} and subsequent evolution via mass transfer~\cite{Stegmann:2020kgb}.

The LVK employs a search that restricts spins to be aligned with the orbital angular momentum, and covers binaries with masses between $2\, M_{\odot}$ and $500\, M_{\odot}$~\cite{KAGRA:2021vkt}. We generated a reduced version of this LVK search by similarly restricting spins to be aligned with the orbital angular momentum, but covering the same NSBH parameter space as our generic spin search. While the full LVK search is also configured to find GW signals observed in one~\cite{Davies:2022thw} or more~\cite{Usman:2015kfa} GW detectors, we limit all searches in this study to find coincident signals in two GW detectors. Since they are the most sensitive, we restrict attention to data collected by LIGO-Hanford and LIGO-Livingston~\cite{LIGOScientific:2014pky}.

Once the search sensitivity has been estimated, the number of confident GW signals found by our search must be determined. A confident GW signal is an event of astrophysical origin with a False Alarm Rate (FAR) smaller than $10^{-2}\, \mathrm{yr}^{-1}$. The False Alarm Rate is defined as how often an event of equivalent or greater significance would occur. To date,the LVK has reported several GW signals from NSBH binaries~\cite{LIGOScientific:2020zkf,KAGRA:2021vkt,LIGOScientific:2021qlt}. However, our search found only one confident detection: (see Figure 1 in Supplementary Material~\citep{Harry:2025aaa}): GW200115\_042309~\cite{LIGOScientific:2021qlt}, a multi-detector NSBH candidate reported by the LVK collaboration~\cite{LIGOScientific:2021qlt,KAGRA:2021vkt,LIGOScientific:2024elc}. We did not find GW200105\_162426~\cite{LIGOScientific:2021qlt} since this was a single detector candidate when ignoring the data collected by Virgo~\cite{acernese2014advanced}. Our search found GW200115\_042309 with a FAR of $1.8\times 10^{-5}\, \mathrm{yr}^{-1}$ and a probability of astrophysical origin $99.9\%$. Through Bayesian inference
techniques, GW200115\_042309 was inferred to have minimal precession; it was found that the effective perpendicular spin $\lesssim 0.4$~\cite{LIGOScientific:2021qlt}.

Once the search sensitivity has been estimated and the number of confident GW observations determined, the merger rate density can be calculated. Given that our generic-spin search did not find any evidence for NSBH binaries with effective perpendicular spin $> 0.4$, we place upper limits on the merger rate of a theoretical subpopulation of precessing NSBHs in the local Universe, see Supplementary Material~\citep{Harry:2025aaa} for details. Defining a precessing subpopulation of NSBH binaries as effective perpendicular spin $> 0.4$, we infer an upper limit on the merger rate at 90\% confidence to be $R_{90} = 79\, \mathrm{Gpc}^{-3}\mathrm{yr}^{-1}$. This reduces to $R_{50} = 13\, \mathrm{Gpc}^{-3}\mathrm{yr}^{-1}$ at 50\% confidence. Our generic-spin search reduces the upper limit on the merger rate by 40\% compared to a comparable search using only aligned-spin waveforms.

Our generic-spin search also allows us to place tighter
constraints on the overall rate of NSBH mergers in the local Universe.
Assuming a Poisson likelihood over the astrophysical rate, a Jeffreys prior, and marginalizing over a 15\% uncertainty arising from calibration uncertainties from the GW detectors, we tighten the merger rate density of NSBHs in the local Universe to $44^{+102}_{-37}\, \mathrm{Gpc}^{-3}\mathrm{yr}^{-1}$; a reduction of $16\%$ on average compared to the merger rate estimated when using a search with no spin-precession (similar to those currently adopted by the LVK)~\cite{Usman:2015kfa} $52^{+130}_{-44}\, \mathrm{Gpc}^{-3}\mathrm{yr}^{-1}$. Although our reported merger rate density assumes a single confident detection and data restricted to O3 only, we expect a similar reduction when more confident detections and data are included. Indeed, if we assume two confident detections, we similarly observe a 16\% reduction in the inferred NSBH merger rate.

In the upper right-hand panel of Figure~\ref{fig:rate}, we show the change in the NSBH rate as a function of the effective perpendicular spin. Assuming negligible spin-precession (effective perpendicular spin $\sim 0$), we find comparable merger rate estimates between our search and an equivalent LVK analysis. This is expected since the change in sensitive volume for NSBHs with negligible spin-precession remains close to 0. However, as soon as spin-precession becomes significant (effective perpendicular spin $> 0$), we infer a smaller merger rate than previously believed. When considering the fully precessing parameter space, we observe a merger rate that is 16\% lower than that estimated with the tools currently adopted by the LVK. We report that the change in the NSBH merger rate is approximately linear in the effective perpendicular spin.

\begin{table*}
\begin{center}
\begin{tabular}{l | c c c c }
\hline
\hline
Name & GPS time & False Alarm Rate ($\mathrm{yr^{-1}}$) & \,\,\,\,SNR\,\,\,\, & $p_{\mathrm{astro}}$ (\%) \\
\hline
GW190421\_203205 & 1239913943.259 & $6.7$ & 9.7 & 0.7 \\
GW190916\_134302 & 1252676600.135 & $10.0$  & 9.5 & 0.5 \\
GW191225\_091958 & 1261300816.100 & $8.3$  & 9.8 & 0.5 \\
GW200310\_030555 & 1267844773.573 & $10.0$  & 9.7 & 0.4 \\
\hline
\hline
\end{tabular}
\caption{Properties of the sub-threshold NSBH candidates uniquely found by our search. For each candidate we report the GPS time, False Alarm Rate, signal-to-noise ratio (SNR), and probability of astrophysical origin $p_{\mathrm{astro}}$. \vspace{-2em} 
}
\label{tab:parameters}
\end{center}
\end{table*}

\textbf{\textit{Impact on the underlying NSBH population.}}—Quantifying the fraction of precessing NSBH binaries in the Universe has significant implications for understanding the formation mechanism of NSBHs. We estimate an upper limit on the fraction of precessing NSBH binaries by directly comparing the rate of precessing binaries (assuming zero observations) with the expected rate of NSBH mergers for the population as a whole. If the two rate estimates are in agreement, zero precessing NSBH binaries are expected in the Universe. In the bottom right-hand panel of Figure.~\ref{fig:rate}, we show the fraction of precessing NSBH binaries as a function of $\chi_{p}$. Assuming an independent subpopulation of precessing NSBH binaries that has $\chi_{p} > 0.4$, we estimate that no more than 30\% of NSBH binaries in the Universe undergo spin-precession. This increases to $35\%$ for $\chi_{\mathrm{p}} > 0.8$. Owing to the improved sensitivity of our generic-spin search, we see a significant reduction in the estimated fraction of precessing NSBH binaries compared to that obtained from an aligned-spin search (similar to those currently adopted by the LVK). Assuming an independent subpopulation of precessing NSBH binaries with $\chi_{p} > 0.8$, ignoring all prior beliefs, we reduce the expected fraction of precessing NSBH binaries in the Universe by $\sim 40\%$. We note that this is an indicative measurement and does not account for the uncertainties on the rate estimates.

\textbf{\textit{New marginal NSBH observations.}}— We report four new sub-threshold NSBH candidates, GW190421\_203205, GW190916\_134302, GW191225\_091958, GW200310\_030555, all with False Alarm Rates $10\,\mathrm{yr}^{-1} < \mathrm{FAR} < 6\, \mathrm{yr}^{-1}$ and probability of astrophysical origin $p_{\mathrm{astro}} < 1\%$. Each candidate's FAR, signal-to-noise ratio and probability of astrophysical origin is given in Table 1. Given the small probability of astrophysical origin, it is likely that these are of terrestrial origin and were therefore not included in our merger rate estimate.

Since our generic-spin search found these marginal candidates, it is likely that they exhibit strong evidence for spin-precession. If real, they are astrophysically significant as they differ from current population estimates for NSBHs~\cite{Biscoveanu:2022iue}, which favours the isolated formation mechanism~\cite{Ye:2019xvf}. We perform Bayesian inference, under the assumption that these sub-threshold NSBH candidates are of astrophysical origin, to infer the properties of each source. The details of our Bayesian analyses are given in Supplementary Material~\citep{Harry:2025aaa}.

Figure~\ref{fig:pe} shows the posterior probability distribution for three of the four candidate's primary mass and effective perpendicular spin. We do not include GW190421\_203205 since its posteriors are highly peaked, indicating a possible issue during sampling. This makes it difficult to inspect the 90\% credible interval; we infer the primary mass to be $8.1^{+0.7}_{-0.2}\, M_{\odot}$ and effective perpendicular spin $0.228^{+0.003}_{-0.003}$. We find that all but one candidate has significant spin-precession, with effective perpendicular spin constrained away from 0, and away from the prior indicating an informative measurement. If real, these newly reported marginal NSBH candidates indicate a distinct population likely formed through dynamic capture.

The inferred secondary masses generally remain consistent with the observed population of neutron stars; we infer $m_{2} = 1.1^{+1.1}_{-0.5}\, M_{\odot}, 1.1^{+0.4}_{-0.2}\, M_{\odot}$ and $1.2^{+0.2}_{-0.2}\, M_{\odot}$ for GW190916\_134302, GW191225\_091958 and GW200310\_030555 respectively. GW190421\_203205 infers a notably smaller secondary mass $m_{2} = 0.791^{+0.003}_{-0.159}\, M_{\odot}$ but this could be a consequence of the sampling issues highlighted above. For all candidates except GW190916\_134302, the effective spin parallel with the orbital angular momentum~\cite{Ajith:2011ec} is negative, with more than $88\%$ probability. This remains consistent with the dynamic capture hypothesis.

\begin{figure}[t!]
  \includegraphics[width=0.45\textwidth]{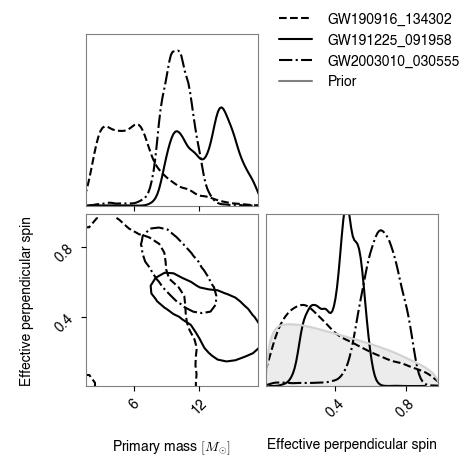}
  \caption{
  Corner plot showing the marginalized posterior probability for the primary mass and effective perpendicular spin for the newly reported sub-threshold NSBH candidates. Contours represent the 90\% credible interval. The prior for the effective perpendicular spin is shown for comparison.
  }
  \label{fig:pe}
\end{figure}

\textbf{\textit{Conclusion.}}— We present for the first time a dedicated template-based matched filter search for NSBHs that is sensitive to the whole NSBH population. We apply our search to all data collected during the third GW observing run, and place upper limits on the
merger rate of a theorised subpopulation of precessing NSBH candidates. We further constrain the overall NSBH merger rate, and show that
on average the inferred NSBH merger rate is reduced by 16\% compared to the results we compute when assuming a search that requires no misalignment between component spin and orbital angular momentum (similar to those currently adopted by the LVK). Although we only considered data collected during O3, we envisage a similar reduction in merger rate when more data are considered. We further identify four new subthreshold NSBH candidates, all with strong imprints of spin-precession. However, we note that these are most likely to be of terrestrial origin.

Given that our more refined NSBH search contains only $\sim 3\times$ more templates than a reduced search currently adopted by the LVK, we highly encourage its use in the future. Our search has the potential to help determine the formation mechanism of NSBHs in the local Universe, as well as constrain properties of the surrounding environment. Moreover, our generic-spin search may also discover precessing NSBHs that would otherwise remain undetected.

While this is the first time that a dedicated NSBH search that incorporates the effects of spin-precession has been performed on Advanced LIGO and Virgo observing run data, we note that our search neglects eccentricity and does not include subdominant GW multipole moments.
It has been shown that it is significantly more important to include precession than subdominant multipoles in NSBH searches~\cite{Dhurkunde:2022aek}, and we further verify this in Supplementary Material~\citep{Harry:2025aaa}. The distribution of eccentricity, from a variety of channels, was shown in~\cite{Dhurkunde:2023qoe}. The authors show that in general the eccentricity of the binary at 10Hz was less than 0.1 for at least 90\% of systems. This level of eccentricity is small enough such that non-eccentric searches would still be able to confidently detect the observed GW signals~\cite{Huerta:2013qb}. Nevertheless, the possible observation that the NSBH GW200105\_162426 has an eccentricity $\sim 0.2$ at 20Hz emission frequency may elevate the priority of developing eccentric search techniques, see for e.g.~\cite{Dhurkunde:2023qoe}.
\\
\par
We thank Gareth Cabourn Davies, Thomas Dent, Rahul Dhurkunde and Francesco Pannarle for comments on this manuscript, and Laura Nuttall for discussion throughout this project. C.H thanks Zoheyr Doctor, Vicky Kalogera and the Center for Interdisciplinary Exploration and Research in Astrophysics (CIERA) for hosting C.H in 2023, and for their guidance on running {\sc{GWPopulation}}. I.H thanks the STFC for support through the grants ST/T000333/1 and ST/V005715/1, and C.H thanks the UKRI Future Leaders Fellowship for support through the grant MR/T01881X/1. We are also grateful for computational resources provided by the LIGO Laboratory and supported by National Science Foundation Grants PHY-0757058 and PHY-0823459, as well as the SCIAMA High Performance Compute (HPC) cluster which is supported by the ICG and the University of Portsmouth. This material is based upon work supported by NSF's LIGO Laboratory, which is a major facility fully funded by the National Science Foundation. This research has made use of data, software and/or web tools obtained from the Gravitational Wave Open Science Center (\href{https://www.gw-openscience.org}{https://www.gw-openscience.org}), a service of LIGO Laboratory, the LIGO Scientific Collaboration and the Virgo Collaboration. LIGO is funded by the U.S. National Science Foundation. Virgo is funded by the French Centre National de Recherche Scientifique (CNRS), the Italian Istituto Nazionale della Fisica Nucleare (INFN) and the Dutch Nikhef, with contributions by Polish and Hungarian institutes.

All GW data analysed in this work can be obtained through the Gravitational Wave Open Science Center (\href{https://www.gw-openscience.org}{https://www.gw-openscience.org}). Details regarding our generic-spin NSBH search code, including template bank generation and running our search of a chunk of O3 data, can be found at~\cite{McIsaac:data_release}. Posterior samples obtained for the marginal NSBH observations detailed in Table.~\ref{tab:parameters} can be found at~\cite{Harry:data_release}.

\bibliographystyle{unsrt}
\bibliography{refs}

\begin{thebibliography}{10}

\bibitem{Usman:2015kfa}
Samantha~A. Usman et~al.
\newblock {The PyCBC search for gravitational waves from compact binary
  coalescence}.
\newblock {\em Class. Quant. Grav.}, 33(21):215004, 2016.

\bibitem{Sachdev:2019vvd}
Surabhi Sachdev et~al.
\newblock {The GstLAL Search Analysis Methods for Compact Binary Mergers in
  Advanced LIGO's Second and Advanced Virgo's First Observing Runs}.
\newblock 1 2019.

\bibitem{Chu:2020pjv}
Qi~Chu et~al.
\newblock {SPIIR online coherent pipeline to search for gravitational waves
  from compact binary coalescences}.
\newblock {\em Phys. Rev. D}, 105(2):024023, 2022.

\bibitem{Aubin:2020goo}
F.~Aubin et~al.
\newblock {The MBTA pipeline for detecting compact binary coalescences in the
  third LIGO\textendash{}Virgo observing run}.
\newblock {\em Class. Quant. Grav.}, 38(9):095004, 2021.

\bibitem{Apostolatos:1994mx}
Theocharis~A. Apostolatos, Curt Cutler, Gerald~J. Sussman, and Kip~S. Thorne.
\newblock {Spin induced orbital precession and its modulation of the
  gravitational wave forms from merging binaries}.
\newblock {\em Phys. Rev. D}, 49:6274--6297, 1994.

\bibitem{Schmidt:2014iyl}
Patricia Schmidt, Frank Ohme, and Mark Hannam.
\newblock {Towards models of gravitational waveforms from generic binaries II:
  Modelling precession effects with a single effective precession parameter}.
\newblock {\em Phys. Rev. D}, 91(2):024043, 2015.

\bibitem{DalCanton:2014qjd}
Tito Dal~Canton, Andrew~P. Lundgren, and Alex~B. Nielsen.
\newblock {Impact of precession on aligned-spin searches for
  neutron-star\textendash{}black-hole binaries}.
\newblock {\em Phys. Rev. D}, 91(6):062010, 2015.

\bibitem{Harry:2016ijz}
Ian Harry, Stephen Privitera, Alejandro Boh\'e, and Alessandra Buonanno.
\newblock {Searching for Gravitational Waves from Compact Binaries with
  Precessing Spins}.
\newblock {\em Phys. Rev. D}, 94(2):024012, 2016.

\bibitem{Burgay:2003jj}
Marta Burgay et~al.
\newblock {An Increased estimate of the merger rate of double neutron stars
  from observations of a highly relativistic system}.
\newblock {\em Nature}, 426:531--533, 2003.

\bibitem{Kalogera:1999tq}
Vassiliki Kalogera.
\newblock {Spin orbit misalignment in close binaries with two compact objects}.
\newblock {\em Astrophys. J.}, 541:319--328, 2000.

\bibitem{Chattopadhyay:2020lff}
Debatri Chattopadhyay, Simon Stevenson, Jarrod~R. Hurley, Matthew Bailes, and
  Floor Broekgaarden.
\newblock {Modelling neutron star{\textendash}black hole binaries: future
  pulsar surveys and gravitational wave detectors}.
\newblock {\em Mon. Not. Roy. Astron. Soc.}, 504(3):3682--3710, 2021.

\bibitem{Broekgaarden:2021iew}
Floor~S. Broekgaarden, Edo Berger, Coenraad~J. Neijssel, Alejandro
  Vigna-G{\'o}mez, Debatri Chattopadhyay, Simon Stevenson, Martyna Chruslinska,
  Stephen Justham, Selma~E. de~Mink, and Ilya Mandel.
\newblock {Impact of massive binary star and cosmic evolution on gravitational
  wave observations I: black hole{\textendash}neutron star mergers}.
\newblock {\em Mon. Not. Roy. Astron. Soc.}, 508(4):5028--5063, 2021.

\bibitem{Rodriguez:2016vmx}
Carl~L. Rodriguez, Michael Zevin, Chris Pankow, Vasilliki Kalogera, and
  Frederic~A. Rasio.
\newblock {Illuminating Black Hole Binary Formation Channels with Spins in
  Advanced LIGO}.
\newblock {\em Astrophys. J. Lett.}, 832(1):L2, 2016.

\bibitem{LIGOScientific:2021qlt}
R.~Abbott et~al.
\newblock {Observation of Gravitational Waves from Two Neutron
  Star\textendash{}Black Hole Coalescences}.
\newblock {\em Astrophys. J. Lett.}, 915(1):L5, 2021.

\bibitem{Morras:2025xfu}
Gonzalo Morras, Geraint Pratten, and Patricia Schmidt.
\newblock {Orbital eccentricity in a neutron star - black hole binary}.
\newblock 3 2025.

\bibitem{KAGRA:2023pio}
R.~Abbott et~al.
\newblock {Open Data from the Third Observing Run of LIGO, Virgo, KAGRA, and
  GEO}.
\newblock {\em Astrophys. J. Suppl.}, 267(2):29, 2023.

\bibitem{Vitale:2014mka}
Salvatore Vitale, Ryan Lynch, John Veitch, Vivien Raymond, and Riccardo
  Sturani.
\newblock {Measuring the spin of black holes in binary systems using
  gravitational waves}.
\newblock {\em Phys. Rev. Lett.}, 112(25):251101, 2014.

\bibitem{Harry:2025aaa}
{See Supplemental Material at
  \href{http://link.aps.org/supplemental/10.1103/cqqn-gl4y}{http://link.aps.org/supplemental/10.1103/cqqn-gl4y}
  for an explanation of our matched filter search, and how we inferred the
  merger rate of NSBH binaries in the local Universe.}

\bibitem{Allen:2005fk}
Bruce Allen, Warren~G. Anderson, Patrick~R. Brady, Duncan~A. Brown, and Jolien
  D.~E. Creighton.
\newblock {FINDCHIRP: An Algorithm for detection of gravitational waves from
  inspiraling compact binaries}.
\newblock {\em Phys. Rev. D}, 85:122006, 2012.

\bibitem{Babak:2012zx}
S.~Babak et~al.
\newblock {Searching for gravitational waves from binary coalescence}.
\newblock {\em Phys. Rev. D}, 87(2):024033, 2013.

\bibitem{Venumadhav:2019tad}
Tejaswi Venumadhav, Barak Zackay, Javier Roulet, Liang Dai, and Matias
  Zaldarriaga.
\newblock {New search pipeline for compact binary mergers: Results for binary
  black holes in the first observing run of Advanced LIGO}.
\newblock {\em Phys. Rev. D}, 100(2):023011, 2019.

\bibitem{McIsaac:2023ijd}
Connor McIsaac, Charlie Hoy, and Ian Harry.
\newblock {Search technique to observe precessing compact binary mergers in the
  advanced detector era}.
\newblock {\em Phys. Rev. D}, 108(12):123016, 2023.

\bibitem{Schmidt:2024jbp}
Stefano Schmidt et~al.
\newblock {Searching for gravitational-wave signals from precessing black hole
  binaries with the GstLAL pipeline}.
\newblock {\em Phys. Rev. D}, 110(2):023038, 2024.

\bibitem{Fairhurst:2019vut}
Stephen Fairhurst, Rhys Green, Charlie Hoy, Mark Hannam, and Alistair Muir.
\newblock {Two-harmonic approximation for gravitational waveforms from
  precessing binaries}.
\newblock {\em Phys. Rev. D}, 102(2):024055, 2020.

\bibitem{Pratten:2020ceb}
Geraint Pratten et~al.
\newblock {Computationally efficient models for the dominant and subdominant
  harmonic modes of precessing binary black holes}.
\newblock {\em Phys. Rev. D}, 103(10):104056, 2021.

\bibitem{Pratten:2020fqn}
Geraint Pratten, Sascha Husa, Cecilio Garcia-Quiros, Marta Colleoni, Antoni
  Ramos-Buades, Hector Estelles, and Rafel Jaume.
\newblock {Setting the cornerstone for a family of models for gravitational
  waves from compact binaries: The dominant harmonic for nonprecessing
  quasicircular black holes}.
\newblock {\em Phys. Rev. D}, 102(6):064001, 2020.

\bibitem{Schmidt:2012rh}
Patricia Schmidt, Mark Hannam, and Sascha Husa.
\newblock {Towards models of gravitational waveforms from generic binaries: A
  simple approximate mapping between precessing and non-precessing inspiral
  signals}.
\newblock {\em Phys. Rev. D}, 86:104063, 2012.

\bibitem{Hamilton:2021pkf}
Eleanor Hamilton, Lionel London, Jonathan~E. Thompson, Edward Fauchon-Jones,
  Mark Hannam, Chinmay Kalaghatgi, Sebastian Khan, Francesco Pannarale, and
  Alex Vano-Vinuales.
\newblock {Model of gravitational waves from precessing black-hole binaries
  through merger and ringdown}.
\newblock {\em Phys. Rev. D}, 104(12):124027, 2021.

\bibitem{Dhurkunde:2022aek}
Rahul Dhurkunde and Alexander~H. Nitz.
\newblock {Sensitivity of spin-aligned searches for neutron star-black hole
  systems using future detectors}.
\newblock {\em Phys. Rev. D}, 106(10):103035, 2022.

\bibitem{LIGOScientific:2016ebi}
B.~P. Abbott et~al.
\newblock {Supplement: The Rate of Binary Black Hole Mergers Inferred from
  Advanced LIGO Observations Surrounding GW150914}.
\newblock {\em Astrophys. J. Suppl.}, 227(2):14, 2016.

\bibitem{metropolis1949monte}
Nicholas Metropolis and Stanislaw Ulam.
\newblock The monte carlo method.
\newblock {\em Journal of the American statistical association},
  44(247):335--341, 1949.

\bibitem{Veitch:2014wba}
J.~Veitch et~al.
\newblock {Parameter estimation for compact binaries with ground-based
  gravitational-wave observations using the LALInference software library}.
\newblock {\em Phys. Rev. D}, 91(4):042003, 2015.

\bibitem{Skilling2004}
John Skilling.
\newblock Nested sampling.
\newblock In {\em {AIP} Conference Proceedings}. {AIP}, 2004.

\bibitem{Skilling:2006gxv}
John Skilling.
\newblock {Nested sampling for general Bayesian computation}.
\newblock {\em Bayesian Analysis}, 1(4):833--859, 2006.

\bibitem{Speagle:2019ivv}
Joshua~S. Speagle.
\newblock {dynesty: a dynamic nested sampling package for estimating Bayesian
  posteriors and evidences}.
\newblock {\em Mon. Not. Roy. Astron. Soc.}, 493(3):3132--3158, 2020.

\bibitem{Ashton:2018jfp}
Gregory Ashton et~al.
\newblock {BILBY: A user-friendly Bayesian inference library for
  gravitational-wave astronomy}.
\newblock {\em Astrophys. J. Suppl.}, 241(2):27, 2019.

\bibitem{Romero-Shaw:2020owr}
I.~M. Romero-Shaw et~al.
\newblock {Bayesian inference for compact binary coalescences with bilby:
  validation and application to the first LIGO\textendash{}Virgo
  gravitational-wave transient catalogue}.
\newblock {\em Mon. Not. Roy. Astron. Soc.}, 499(3):3295--3319, 2020.

\bibitem{KAGRA:2021vkt}
R.~Abbott et~al.
\newblock {GWTC-3: Compact Binary Coalescences Observed by LIGO and Virgo
  during the Second Part of the Third Observing Run}.
\newblock {\em Phys. Rev. X}, 13(4):041039, 2023.

\bibitem{Morisaki:2021ngj}
Soichiro Morisaki.
\newblock {Accelerating parameter estimation of gravitational waves from
  compact binary coalescence using adaptive frequency resolutions}.
\newblock {\em Phys. Rev. D}, 104(4):044062, 2021.

\bibitem{Thompson:2020nei}
Jonathan~E. Thompson, Edward Fauchon-Jones, Sebastian Khan, Elisa Nitoglia,
  Francesco Pannarale, Tim Dietrich, and Mark Hannam.
\newblock {Modeling the gravitational wave signature of neutron star black hole
  coalescences}.
\newblock {\em Phys. Rev. D}, 101(12):124059, 2020.

\bibitem{Matas:2020wab}
Andrew Matas et~al.
\newblock {Aligned-spin neutron-star\textendash{}black-hole waveform model
  based on the effective-one-body approach and numerical-relativity
  simulations}.
\newblock {\em Phys. Rev. D}, 102(4):043023, 2020.

\bibitem{Gonzalez:2022prs}
Alejandra Gonzalez, Rossella Gamba, Matteo Breschi, Francesco Zappa, Gregorio
  Carullo, Sebastiano Bernuzzi, and Alessandro Nagar.
\newblock {Numerical-relativity-informed effective-one-body model for
  black-hole\textendash{}neutron-star mergers with higher modes and spin
  precession}.
\newblock {\em Phys. Rev. D}, 107(8):084026, 2023.

\bibitem{LIGOScientific:2024elc}
A.~G. Abac et~al.
\newblock {Observation of Gravitational Waves from the Coalescence of a
  2.5\textendash{}4.5 $M_{\odot}$ Compact Object and a Neutron Star}.
\newblock {\em Astrophys. J. Lett.}, 970(2):L34, 2024.

\bibitem{Kumar:2016zlj}
Prayush Kumar, Michael P\"urrer, and Harald~P. Pfeiffer.
\newblock {Measuring neutron star tidal deformability with Advanced LIGO: a
  Bayesian analysis of neutron star - black hole binary observations}.
\newblock {\em Phys. Rev. D}, 95(4):044039, 2017.

\bibitem{Cornish:2014kda}
Neil~J. Cornish and Tyson~B. Littenberg.
\newblock {BayesWave: Bayesian Inference for Gravitational Wave Bursts and
  Instrument Glitches}.
\newblock {\em Class. Quant. Grav.}, 32(13):135012, 2015.

\bibitem{Farr:calnote}
W.~M. Farr, B.~Farr, and Littenberg T.
\newblock Modelling calibration errors in cbc waveforms.
\newblock DCC, October 2014.

\bibitem{ligo_scientific_collaboration_and_virgo_2021_5546663}
LIGO~Scientific Collaboration, Virgo Collaboration, and KAGRA Collaboration.
\newblock Gwtc-3: Compact binary coalescences observed by ligo and virgo during
  the second part of the third observing run — parameter estimation data
  release, November 2021.

\bibitem{Mandel:2018mve}
Ilya Mandel, Will~M. Farr, and Jonathan~R. Gair.
\newblock {Extracting distribution parameters from multiple uncertain
  observations with selection biases}.
\newblock {\em Mon. Not. Roy. Astron. Soc.}, 486(1):1086--1093, 2019.

\bibitem{Farr:2019rap}
Will~M. Farr.
\newblock {Accuracy Requirements for Empirically-Measured Selection Functions}.
\newblock {\em Research Notes of the AAS}, 3(5):66, 2019.

\bibitem{Tiwari:2017ndi}
Vaibhav Tiwari.
\newblock {Estimation of the Sensitive Volume for Gravitational-wave Source
  Populations Using Weighted Monte Carlo Integration}.
\newblock {\em Class. Quant. Grav.}, 35(14):145009, 2018.

\bibitem{2019PhRvD.100d3030T}
Colm {Talbot}, Rory {Smith}, Eric {Thrane}, and Gregory~B. {Poole}.
\newblock {Parallelized inference for gravitational-wave astronomy}.
\newblock {\em Phys. Rev. D}, 100(4):043030, August 2019.

\bibitem{Biscoveanu:2022iue}
Sylvia Biscoveanu, Philippe Landry, and Salvatore Vitale.
\newblock {Population properties and multimessenger prospects of neutron
  star\textendash{}black hole mergers following GWTC-3}.
\newblock {\em Mon. Not. Roy. Astron. Soc.}, 518(4):5298--5312, 2022.

\bibitem{Chatziioannou:2020pqz}
Katerina Chatziioannou.
\newblock {Neutron star tidal deformability and equation of state constraints}.
\newblock {\em Gen. Rel. Grav.}, 52(11):109, 2020.

\bibitem{Damour:2009vw}
Thibault Damour and Alessandro Nagar.
\newblock {Relativistic tidal properties of neutron stars}.
\newblock {\em Phys. Rev. D}, 80:084035, 2009.

\bibitem{Binnington:2009bb}
Taylor Binnington and Eric Poisson.
\newblock {Relativistic theory of tidal Love numbers}.
\newblock {\em Phys. Rev. D}, 80:084018, 2009.

\bibitem{Foucart:2013psa}
Francois Foucart, Luisa Buchman, Matthew~D. Duez, Michael Grudich, Lawrence~E.
  Kidder, Ilana MacDonald, Abdul Mroue, Harald~P. Pfeiffer, Mark~A. Scheel, and
  Bela Szilagyi.
\newblock {First direct comparison of nondisrupting neutron star-black hole and
  binary black hole merger simulations}.
\newblock {\em Phys. Rev. D}, 88(6):064017, 2013.

\bibitem{Pannarale:2011pk}
Francesco Pannarale, Luciano Rezzolla, Frank Ohme, and Jocelyn~S. Read.
\newblock {Will black hole-neutron star binary inspirals tell us about the
  neutron star equation of state?}
\newblock {\em Phys. Rev. D}, 84:104017, 2011.

\bibitem{KAGRA:2021duu}
R.~Abbott et~al.
\newblock {Population of Merging Compact Binaries Inferred Using Gravitational
  Waves through GWTC-3}.
\newblock {\em Phys. Rev. X}, 13(1):011048, 2023.

\bibitem{Hannam:2021pit}
Mark Hannam et~al.
\newblock {General-relativistic precession in a black-hole binary}.
\newblock {\em Nature}, 610(7933):652--655, 2022.

\bibitem{Valsecchi:2010cw}
Francesca Valsecchi, Evert Glebbeek, Will~M. Farr, Tassos Fragos, Bart Willems,
  Jerome~A. Orosz, Jifeng Liu, and Vassiliki Kalogera.
\newblock {Formation of the black-hole binary M33 X-7 via mass-exchange in a
  tight massive system}.
\newblock {\em Nature}, 468:77, 2010.

\bibitem{Wong:2011eg}
Tsing-Wai Wong, Francesca Valsecchi, Tassos Fragos, and Vassiliki Kalogera.
\newblock {Understanding Compact Object Formation and Natal Kicks. III. The
  case of Cygnus X-1}.
\newblock {\em Astrophys. J.}, 747:111, 2012.

\bibitem{Zhao:2021bhj}
Xueshan Zhao, Lijun Gou, Yanting Dong, Xueying Zheng, James~F. Steiner, James
  C.~A. Miller-Jones, Arash Bahramian, Jerome~A. Orosz, and Ye~Feng.
\newblock {Re-estimating the Spin Parameter of the Black Hole in Cygnus X-1}.
\newblock {\em Astrophys. J.}, 908(2):117, 2021.

\bibitem{Belczynski:2011bv}
Krzysztof Belczynski, Tomasz Bulik, and Charles Bailyn.
\newblock {The fate of Cyg X-1: an empirical lower limit on BH-NS merger rate}.
\newblock {\em Astrophys. J. Lett.}, 742:L2, 2011.

\bibitem{Steinle:2020xej}
Nathan Steinle and Michael Kesden.
\newblock {Pathways for producing binary black holes with large misaligned
  spins in the isolated formation channel}.
\newblock {\em Phys. Rev. D}, 103(6):063032, 2021.

\bibitem{Gerosa:2018wbw}
Davide Gerosa, Emanuele Berti, Richard O'Shaughnessy, Krzysztof Belczynski,
  Michael Kesden, Daniel Wysocki, and Wojciech Gladysz.
\newblock {Spin orientations of merging black holes formed from the evolution
  of stellar binaries}.
\newblock {\em Phys. Rev. D}, 98(8):084036, 2018.

\bibitem{Stegmann:2020kgb}
Jakob Stegmann and Fabio Antonini.
\newblock {Flipping spins in mass transferring binaries and origin of
  spin-orbit misalignment in binary black holes}.
\newblock {\em Phys. Rev. D}, 103(6):063007, 2021.

\bibitem{Davies:2022thw}
Gareth S.~Cabourn Davies and Ian~W. Harry.
\newblock {Establishing significance of gravitational-wave signals from a
  single observatory in the PyCBC offline search}.
\newblock {\em Class. Quant. Grav.}, 39(21):215012, 2022.

\bibitem{LIGOScientific:2014pky}
J.~Aasi et~al.
\newblock {Advanced LIGO}.
\newblock {\em Class. Quant. Grav.}, 32:074001, 2015.

\bibitem{LIGOScientific:2020zkf}
R.~Abbott et~al.
\newblock {GW190814: Gravitational Waves from the Coalescence of a 23 Solar
  Mass Black Hole with a 2.6 Solar Mass Compact Object}.
\newblock {\em Astrophys. J. Lett.}, 896(2):L44, 2020.

\bibitem{acernese2014advanced}
F.~Acernese et~al.
\newblock {Advanced Virgo: a second-generation interferometric gravitational
  wave detector}.
\newblock {\em Class. Quant. Grav.}, 32(2):024001, 2015.

\bibitem{Ye:2019xvf}
Claire~S. Ye, Wen-fai Fong, Kyle Kremer, Carl~L. Rodriguez, Sourav Chatterjee,
  Giacomo Fragione, and Frederic~A. Rasio.
\newblock {On the Rate of Neutron Star Binary Mergers from Globular Clusters}.
\newblock {\em Astrophys. J. Lett.}, 888(1):L10, 2020.

\bibitem{Ajith:2011ec}
P.~Ajith.
\newblock {Addressing the spin question in gravitational-wave searches:
  Waveform templates for inspiralling compact binaries with nonprecessing
  spins}.
\newblock {\em Phys. Rev. D}, 84:084037, 2011.

\bibitem{Dhurkunde:2023qoe}
Rahul Dhurkunde and Alexander~H. Nitz.
\newblock {Search for eccentric NSBH and BNS mergers in the third observing run
  of Advanced LIGO and Virgo}.
\newblock {\em Phys. Rev. D}, 111(10):103018, 2025.

\bibitem{Huerta:2013qb}
E.~A. Huerta and Duncan~A. Brown.
\newblock {Effect of eccentricity on binary neutron star searches in Advanced
  LIGO}.
\newblock {\em Phys. Rev. D}, 87(12):127501, 2013.

\bibitem{McIsaac:data_release}
Connor McIsaac, Charlie Hoy, and Ian Harry.
\newblock {A search technique to observe precessing compact binary mergers in
  the advanced detector era - Data Release}, 2023,
  \href{https://icg-gravwaves.github.io/precessing_search_paper/}{https://icg-gravwaves.github.io/precessing\_search\_paper/}.

\bibitem{Harry:data_release}
Ian Harry and Charlie Hoy.
\newblock {Constraining the neutron star-black hole merger rate - Data
  Release}, 2025,
  \href{https://icg-gravwaves.github.io/nsbh_search_on_O3_paper/}{https://icg-gravwaves.github.io/nsbh\_search\_on\_O3\_paper/}.

\end{thebibliography}
\end{document}